# Structure of a Parabolic Partial Differential Equation on Graphs and Digital spaces. Solution of PDE on Digital Spaces: a Klein Bottle, a Projective Plane, a 4D Sphere and a Moebius Band

Alexander V. Evako

"Dianet", Laboratory of Digital Technologies,  125080 Moscow, Russia

Email address:
evakoa@mail.ru

**Abstract:** This paper studies the structure of a parabolic partial differential equation on graphs and digital n-dimensional manifolds, which are digital models of continuous n-manifolds. Conditions for the existence of solutions of equations are determined and investigated. Numerical solutions of the equation on a Klein bottle, a projective plane, a 4D sphere and a Moebius strip are presented.

## 1. Introduction

In the past decades, non-orientable surfaces such as a Moebius strip, Klein bottle and projective plane have attracted many scientists from different fields. The study was derived from obvious practical and science background. In physics, a considerable interest has emerged in studying lattice models on non-orientable surfaces as new challenging unsolved lattice-statistical problems and as a realization and testing of predictions of the conformal field theory (see, e.g., [13]). In a joint Russian-French-German project a Moebius strip was proposed as a basic element of an airplane wing. Many important technical and physical properties of Moebius-type structural elements can be described by solutions of partial differential equations (PDE), where a Moebius strip serves as a domain. A problem exists in description of electronic and nuclear motions in nano-technology structures and biological networks. Modeling blood flow through a capillary network or road traffic requires a system of differential equations on a graph. Since analytic solutions of PDE can be obtained only in simple geometric regions, for practical problems, it is more reasonable to use computational or numerical solutions. We can do this by implementing as domains graphs and digital spaces, which are discreet counterparts of continuous spaces, and by transferring PDE from a continuous area into discrete one. A review of works devoted to partial differential equations on graphs can be found in [2], [14], and [17]. A serious problem arises because in most of cases, the grid is not a correct counterpart of the continuous area in terms of digital topology and, therefore, cannot properly model the continuous domain. Distinctions between the differential equations on discrete and continuous spaces are also essential. One of differences is stipulated by the fact that a digital space can have just a few points. Another serious difference is linked to the existence of the natural least length in a digital space, defined by the length of the edge connecting two adjacent points of the space. In application to wave processes it means a lack of indefinitely short waves and indefinitely high frequencies that is the lack of the factors frequently conducting to divergences.

In order to build mathematically correct grids it is reasonable to use digital topology methods.

Digital spaces are studied in the framework of digital topology, which plays an important role in analyzing n-dimensional digitized images arising in computer graphics as well as in many areas of science including neuroscience, medical imaging, computer graphics, geoscience and fluid dynamics. Usually, digital objects are represented by graphs whose edges define nearness and connectivity (see, e.g., [3] and [6]). The important feature of an n-surface is a similarity of its properties with properties of its continuous counterpart in terms of algebraic topology. For example, the Euler characteristics and the homology groups of digital n-spheres, a Moebius strip and a Klein bottle are the same as ones of their continuous counterparts ([10] and [11]). In recent years, there has been a considerable amount of works devoted to building two, three and n-dimensional discretization schemes and digital images. In papers [4] and [9], discretization schemes are defined and studied that allow to build digital models of 2-dimensional continuous objects with the same topological properties as their continuous counterparts.

Section 2: The material to be presented below begins with a short description digital spaces and digital n-surfaces studied in [5], [6], [8] such as digital n-dimensional spheres, a digital torus, a digital Klein bottle, etc.

Section 3 defines parabolic differential equations on a graph, a digital space and a network and studies some its properties.

Section 4 presents a numerical solution of a parabolic equation on a digital Klein bottle, a digital projective plane, a digital 4D sphere, a digital Moebius band and a directed network installed in digital space (which, for example, can model a capillary network for blood flow).

# 2. Digital N-surfaces

There are a considerable amount of literature devoted to the study of different approaches to digital lines, surfaces and spaces in the framework of digital topology. Digital topology studies topological properties of discrete objects which are obtained digitizing continuous objects.

This section includes some results related to digital spaces. Traditionally, a digital image has a graph structure (see [3] and [7]). A digital space G is a simple undirected graph G=(V, W) where V=($v_1$, $v_2$, ... $v_n$, …) is a finite or countable set of points, and W = (($v_p v_q$),....) is a set of edges. Topological properties of G as a digital space in terms of adjacency, connectedness and dimensionality are completely defined by set W. Let G and v be a graph and a point of G. In [7], the subgraph O(v) containing all neighbors of v (without v) is called the rim of point v in G. The subgraph U(v)=v∪O(v) containing O(v) as well as point v is called the ball of point v in G. Let (vu) be an edge of G. The subgraph O(vu)=O(v)∩O(u) is called the rim of (vu).

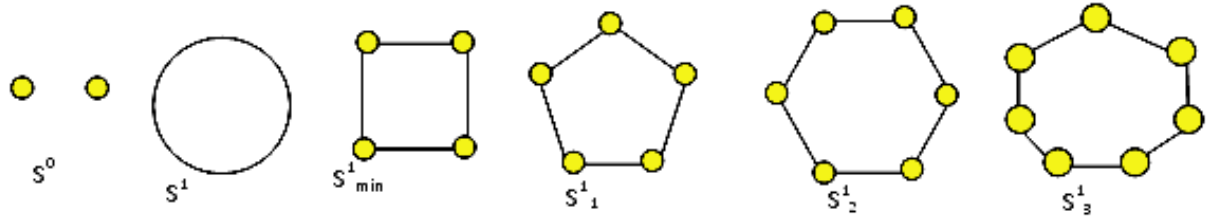

**Fig. 1.** $S^0$ is a digital zero-dimensional sphere, $S^1_{min}$, $S^1_1$, $S^1_2$, $S^1_3$ are digital one-dimensional spheres.

For two graphs G=(X, U) and H=(Y, W) with disjoint point sets X and Y, their join G⊕H is the graph that contains G, H and edges joining every point in G with every point in H.

Contractible graphs are basic elements in this approach.

**Definition 2.1**

A one-point graph is contractible. If G is a contractible graph and H is a contractible subgraph of G then G can be converted into H by sequential deleting simple points.

A point v in graph G is simple if the rim O(v) of v is a contractible graph.

An edge (uv) of a graph G is called simple if the rim O(vu)=O(v)∩O(u) of (uv) is a contractible graph.

By construction, a contractible graph is connected. It follows from definition 1 that a contractible graph can be converted to a point by sequential deleting simple points.

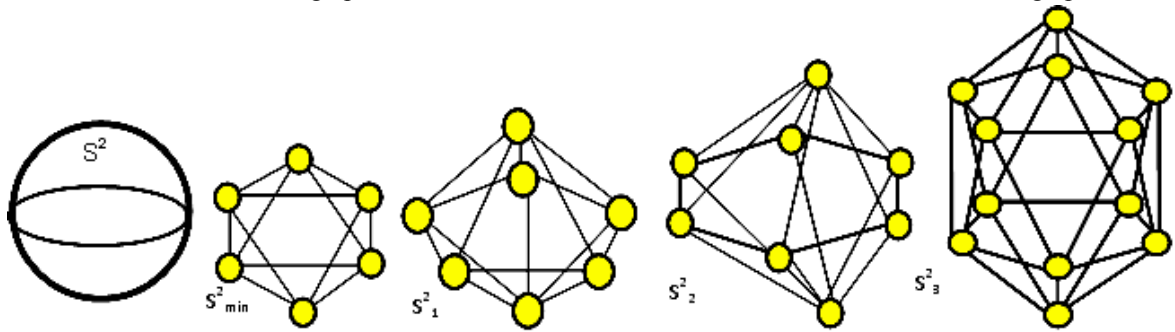

**Fig. 2.** Two-dimensional spheres with a different number of point, any of spheres can be converted into the minimal sphere $S^2_{min}$ by contractible transformations.

**Definition 2.2**

Deletions and attachments of simple points and edges are called contractible transformations.

Graphs G and H are called homotopy equivalent if one of them can be converted to the other one by a sequence of contractible transformations.

Homotopy is an equivalence relation among graphs. Contractible transformations of graphs seem to play the same role in this approach as a homotopy in algebraic topology. In papers [10] and [11], it was shown that contractible transformations retain the Euler characteristic and homology groups of a graph. A digital n-manifold is a special case of a digital n-surface defined and investigated in [6].

**Definition 2.3**

The digital 0-dimensional surface $S^0$(a, b) is a disconnected graph with just two points a and b. For n>0, a digital n-dimensional surface $G^n$ is a nonempty connected graph such that, for each point v of $G^n$, O(v) is a finite digital (n-1)-dimensional surface.

A connected digital n-dimensional surface $G^n$ is called a digital n-sphere, n>0, if for any point v∈$G^n$, the rim O(v) is an (n-1)-sphere and the space $G^n$-v is a contractible graph.

A digital n-dimensional surface $G^n$ is a digital n-manifold if for each point v of $G^n$, O(v) is a finite digital (n-1)-dimensional sphere.

Digital n-manifolds are called homeomorphic if they are homotopy equivalent. A digital n-manifold M can be converted to a homeomorphic digital n-manifold N with the minimal number of points by contractible transformations.

***Definition 2.4***

Let M be a digital n-sphere, n>0, and v be a point belonging to M. The space N=M-v is called *a digital n-disk* with the boundary ∂N=O(v) and the interior IntN=N-∂N.

Let M be an n-manifold and a point v belong to M. Then the space N=M-v is called an n-manifold with the spherical boundary ∂N=O(v) and the interior IntN=N-∂N.

According to definition 2.3, a digital n-disk is a contractible graph. A digital n-disk is a digital counterpart of a continuous n-dimensional disk in Euclidean $E^n$.

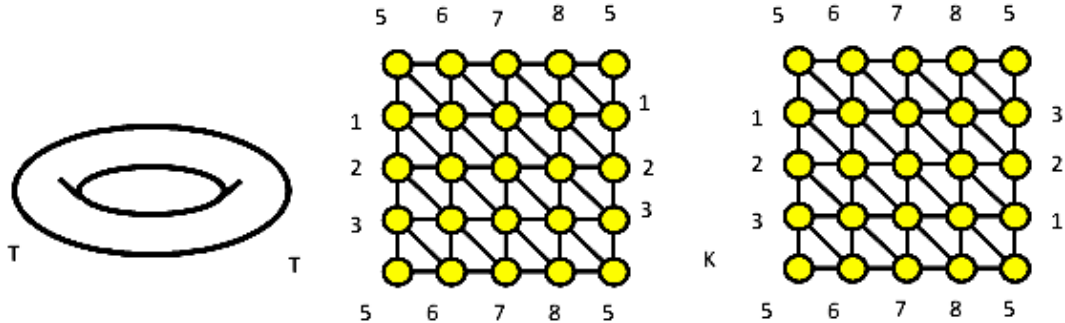

***Fig. 3.*** *Digital 2-dimensional torus T and Klein bottle K with sixteen points.*

The following results were obtained in [6] and [8].

***Theorem 2.1***

The join $S^n_{min}=S^0_1 \oplus S^0_2 \oplus \ldots S^0_{n+1}$ of (n+1) copies of the zero-dimensional sphere $S^0$ is a minimal n-sphere.

Let *M* and *N* be *n* and *m*-spheres. Then $M \oplus N$ is an (n+m+1)-sphere.

Any n-sphere M can be converted to the minimal n-sphere $S_{min}$ by contractible transformations.

Let M be an n-manifold, G and H be contractible subspaces of M and v be a point in M. Then subspaces M-G, M-H and

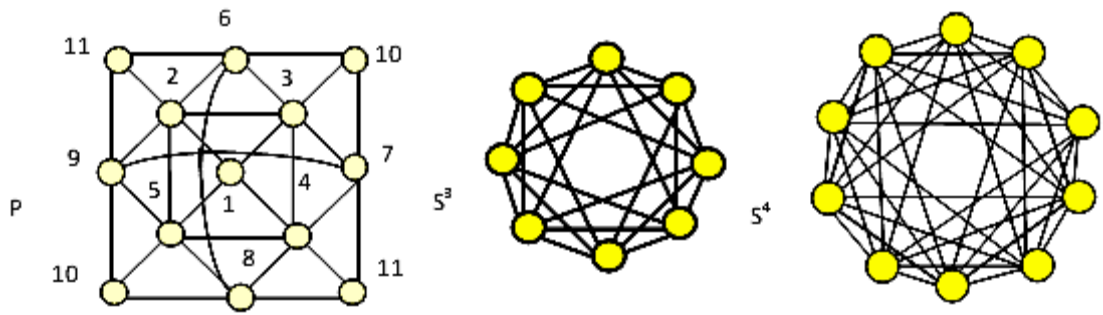

***Fig. 4.*** *P, $S^3$ and $S^4$ are digital two-dimensional projective plane, there- and four-dimensional spheres respectively.*

M-v are all homotopy equivalent to each other. The replacement of an edge with a point increases the number of points in a digital n-manifold.

***Definition 2.5***

Let M be an n-manifold, v and u be adjacent points in M and (vu) be the edge in M. Glue a point x to M in such a way that $O(x)=v \oplus u \oplus O(vu)$, and delete the edge (vu) from the space. This pair of contractible transformations is called the replacement of an edge with a point or R-transformation, R: M→N. The obtained space N is denoted by N=RM=(M∪x)-(vu).

***Theorem 2.2 ([8])***

Let M be an n-manifold and N=RM be a space obtained from M by an R-transformation. Then N is homeomorphic to M.

An R-transformation is a digital homeomorphism because it retains the dimension and other local and global topological features of an n-manifold. R-Transformations increase the number of points in a given n-manifold M retaining the global topology (the homotopy type of M) and the local topology (the homotopy type and the dimension of the neighborhood of any point). A close connection between continuous and digital n-manifolds for n=2 was shown in [4].

A digital 0-dimensional surface is a digital 0-dimensional sphere. Fig. 1 depicts digital zero and one-dimensional spheres.

Fig. 2 shows digital 2-dimensional spheres. All spheres are homeomorphic and can be converted into the minimal sphere $S^2_{min}$ by contractible transformations. Digital torus T and a digital 2-dimensional Klein bottle K are shown in fig. 3. Fig. 4 depicts a digital projective plane P and digital three and four-dimensional spheres $S^3$ and $S^4$ respectively. In the finite difference method for solving partial differential equations in two and three dimensions, a two or three-dimensional continuous domain is replaced by a grid. This grid has to be a digital model of a continuous space. However, in most of cases, the grid is not a correct two or three- dimensional space in terms of digital n-surfaces. For example, consider a standard two-dimensional grid

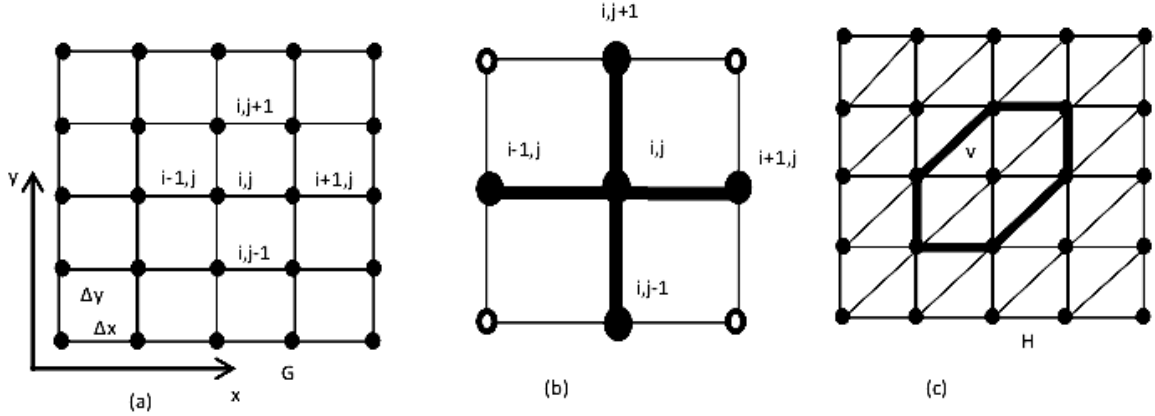

**Fig. 5.** *(a) Finite difference 2D grid for two independent variable x and y, (b) The spatial stencil for the parabolic partial differential equation with two spatial independent variables x and y, (c) Digital 2-dimensional plane.*

G (fig. 5(a)) often used in finite-difference schemes. As one can see, the neighborhood O(v) of any point v consists of four non-adjacent points and, therefore, is not a one-dimensional sphere. Hence, G is not a part of a digital two-dimensional plane, but rather can be seen as a collection of one-dimensional segments. Grid H in fig. 5(c) is a part of a digital plane because the rim O(v) of any interior point v is a digital 1-sphere $S^1$ containing six points and shown in fig. 1. Thus, H is a correct grid which should be used in finite-difference schemes.

## 3. Parabolic PDE on a Graph and a Digital Space

### 3.1. The Structure of Parabolic Partial Differential Equations on a Digital Space

Some results of the investigation of partial differential equations on graphs and digital spaces were published in [5]. Finite difference approximations of the PDE are based upon replacing partial differential equations by finite difference equations using Taylor approximations [15]. Consider a parabolic PDE with two spatial independent variables.

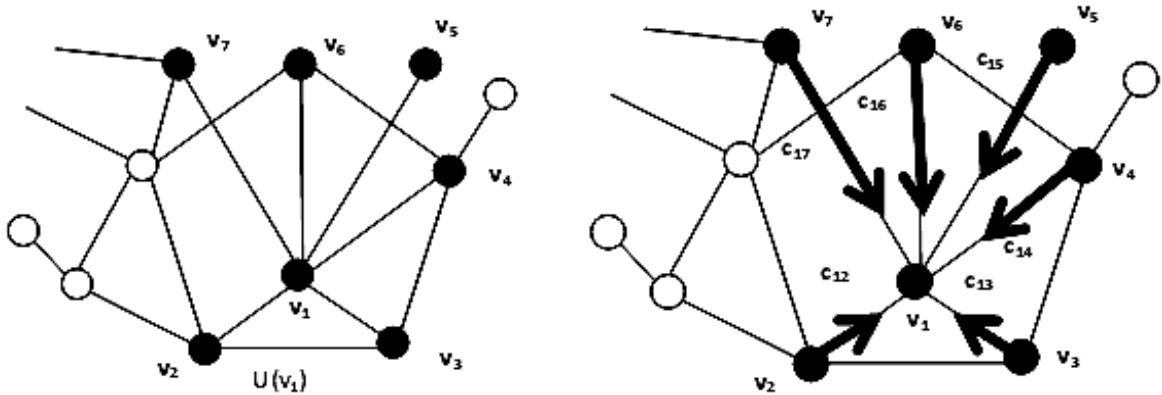

**Fig. 6.** *The ball $U(v_1)$ of point $v_1$ (black points). $F_1^{t+1} = c^t_{11}f^t_1 + c^t_{12}f^t_2 + c^t_{13}f^t_3 + c^t_{14}f^t_4 + c^t_{15}f^t_5 + c^t_{16}f^t_6 + c^t_{17}f^t_7$.*

$$\frac{\partial f}{\partial t} = a\frac{\partial^2 f}{\partial x^2} + b\frac{\partial^2 f}{\partial y^2} + g \qquad (1)$$

where $f = f(x,y,t), a = a(x,y,t), b = b(x,y,t), g = g(x,y,t)$. Using a two-dimensional spatial orthogonal grid G shown in fig. 5(a), the central difference formula for the second derivatives with respect to x and y, and the forward difference formula for the derivative with respect to t, we obtain the following equivalent finite deference equation

$$\frac{f_{i,j}^{n+1}-f_{i,j}^n}{\Delta t} = a_{i,j}^n \frac{f_{i-1,j}^n - 2f_{i,j}^n + f_{i+1,j}^n}{\Delta x^2} + b_{i,j}^n \frac{f_{i,j-1}^n - 2f_{i,j}^n + f_{i,j+1}^n}{\Delta y^2} + g_{i,j}^n \quad (2)$$

where $x = i\Delta x, i = 1,2, \ldots, y = j\Delta y, j = 1,2, \ldots, t = n\Delta t, n = 1,2, \ldots$. Equation (2) can be written as

$$f_{i,j}^{n+1} = e_{i-1,j}^n f_{i-1,j}^n + e_{i-1,j}^n f_{i-1,j}^n + e_{i+1,j}^n f_{i+1,j}^n + e_{i,j-1}^n f_{i,j-1}^n + e_{i,j+1}^n f_{i,j+1}^n + e_{i,j}^n f_{i,j}^n + \Delta t g_{i,j}^n \quad (3)$$

where $e_{i,j}^n = 1 - e_{i-1,j}^n - e_{i+1,j}^n - e_{i,j-1}^n - e_{i,j+1}^n$, $e_{i-1,j}^n = e_{i+1,j}^n = \frac{\Delta t}{\Delta x^2} a_{i,j}^n$, $e_{i,j+1}^n = e_{i,j+1}^n = \frac{\Delta t}{\Delta x^2} b_{i,j}^n$.

Grid G in fig. 5(a) is a graph with points $v_{ps} = (p\Delta x, s\Delta y)$. Point $v_{ij}$ is adjacent to points $v_{i,j\pm 1}$ and $v_{i\pm 1,j}$. Notice that points $v_{ij}, v_{i,j\pm 1}$ and $v_{i\pm 1,j}$ form the ball 1 $U(v_{ij}) = v_{ij} \cup O(v_{ij})$ depicted in fig.5(b). Based on this consideration and equation (3), we can define a dynamic system on a graph G by the set of equations on G.

$$f_p^{t+1} = \sum_{v_k \in U(v_p)} c_{pk}^t f_k^t + g_p^t, p = 1, \ldots n, \text{where } \sum_{v_k \in U(v_p)} c_{pk}^t = 1, p = 1, \ldots n \quad (4)$$

The summation is produced over all points $v_k$ belonging to the ball $U(v_p)$ of point $v_p$. Here $f_k^t$ is the value of the function $f(v_k, t)$ at point $v_k$ of $G$ at the moment $t$, coefficients $c_{pk}^t$ are functions on the pairs of point $(v_p, v_k)$ and $t$ (with domain $V \times V \times t$). If points $v_p$ and $v_k$ are not adjacent, then $c_{pk}^t = 0$.

**Remark 3.1**

In general, the condition $\sum_{v_k \in U(v_p)} c_{pk}^t = 1, p = 1,2, \ldots n$, is not necessary for the PDE. For example, it does not hold for the diffusion equation on a directed network. If all $g_p^t = 0$, then the equation is called homogeneous. Later on in this paper, all $g_p^t = 0$.

**Definition 3.1**

Let G(V, W) be a graph (digital space) with the set of points V=($v_1, v_2,\ldots v_n$), the set of edges W = (($v_p v_q$),….), and $U(v_p)$ be the ball of point $v_p$. A parabolic PDE on G is the set of n equations of the form

$$f_p^{t+1} = \sum_{v_k \in U(v_p)} c_{pk}^t f_k^t, p = 1, \ldots n \quad (5)$$

Thus a PDE on a given point $v_k$ depends on the values of a function on $v_k$ and the points adjacent to $v_k$. Equation (5) is illustrated in fig. 6. The ball $U(v_1)$ consists of black points, and $f_1^{t+1} = c_{11}^t f_1^t + c_{12}^t f_2^t + c_{13}^t f_3^t + c_{14}^t f_4^t + c_{15}^t f_5^t + c_{16}^t f_6^t + c_{17}^t f_7^t$. Since $c_{pk}^t = 0$ if points $v_p$ and $v_k$ are non-adjacent, then set (5) can be written in the form

$$f_p^{t+1} = \sum_{v_k \in U(v_p)} c_{pk}^t f_k^t = \sum_{k=1}^n c_{pk}^t f_k^t, p = 1,2, \ldots n, \quad (6)$$

Equations (6) do not depend explicitly on the dimension of G, if G is a digital space, and can be applied to a graph or a digital space of any dimension or a network. All dimensional features are contained in the local and global structure of G. Set (6) is similar to the set of differential equations on a graph investigated by A. I. Volpert in [16]. Equation (6) can be presented in the matrix form

$$f^t = \begin{bmatrix} f_1^t \\ \cdot \\ f_n^t \end{bmatrix}, C(t) = \begin{bmatrix} c_{11}^t & c_{12}^t & \cdot \\ c_{21}^t & \cdot & \cdot \\ \cdot & \cdot & c_{nn}^t \end{bmatrix}, \quad (7)$$

$$f^{t+1} = C(t) f^t \quad (8)$$

Consider now the initial conditions for solving equation (5). Initial conditions are defined in a regular way by the set of equations

$$f_p^0 = f(v_p, 0), p = 1,2, \ldots n \quad (9)$$

**Definition 3.2**

Equations (5) along with initial conditions (9) are called the initial value problem for the parabolic PDE on a graph $G(v_1, v_2, \ldots v_n)$.

$$f_p^{t+1} = \sum_{v_k \in U(v_p)} c_{pk}^t f_k^t, f_p^0 = f(v_p, 0), p = 1,2, \ldots n \quad (10)$$

Boundary conditions can also be set the usual way. Boundary conditions are affected by what happens at the subgraph H of G. Let H be a subspace of G. Let the value of the function $f(v_k, t)$ at points $v_k \in H$ at the moment $t$ be given by the set

$$f(v_k, t) = f_k^t = s_k^t, v_k \epsilon H \qquad (11)$$

The boundary-value problem on G can be formulated as follows

***Definition 3.3***

Let $G(v_1, v_2, ... v_n)$ be a graph and the set (5) be a differential parabolic equation on G. Let H be a subgraph of G, and the value of the function $f(v_k, t)$ at points $v_k \epsilon H$ at the moment $t$ be defined by boundary conditions (11). Equation (5) along with boundary conditions (11) is called the boundary value problem for the parabolic PDE on a graph G.

$$f_p^{t+1} = \sum_{v_k \in U(v_p)} c_{pk}^t f_k^t, p = 1,2, ... n, f(v_k, t) = s_k^t, v_k \epsilon H, \quad (12)$$

Consider the stability of equation (5) using a standard approach. For $f^t$, the norm is defined as $\|f^t\| = |f_1^t| + ... |f_n^t|$. Equation (5) is called stable according to initial values if there exists a positive b such that $\|f^t\| \leq b\|f^0\|$. for all t.

***Theorem 3.1***

An equation (10) on a graph $G(v_1, v_2, ... v_n)$ is stable according to initial values, i.e., $\|f^t\| \leq M\|f^0\|$, if for any t there is $b, 0 < b < 1/n$, such that $|c_{pk}^t| \leq b, p, k = 1, ... n$.

*Proof*

Consider the equation in form (6). $|f_p^{t+1}| = |\sum_{k=1}^n c_{pk}^t f_k^t| \leq \sum_{k=1}^n |c_{pk}^t||f_k^t| \leq b \sum_{k=1}^n |f_k^t| = b\|f^t\|.$ Then $\|f^{t+1}\| = |f_1^{t+1}| + ... |f_n^{t+1}| \leq b\|f^t\| + ... b\|f^t\| = bn\|f^t\| < 1/n \, n\|f^t\| = \|f^t\|$. Since $\|f^{t+1}\| < \|f^t\|$ for any t, then $\|f^t\| < \|f^0\|$. It completes the proof. □

## 3.2. The Heat and Diffusion Equations

Later in this section we consider homogeneous equations. Diffusion is the process by which particles are transported from one point of a graph $G(v_1, v_2, ... v_n)$ to another points adjacent to the first one as a result of random motions: the number $c_{pk}^t f_k^t$ of particles on point $v_k$ will jump to point $v_p$, the number $c_{pp}^t f_p^t$ of particles on point $v_p$ will stay on $v_p$. Conservation of the total number of particles is obvious, i.e., $f_k^t = \sum_{v_p \in U(v_k)} c_{pk}^t f_k^t$. This means that

$$\sum_{v_p \in U(v_k)} c_{pk}^t = 1, c_{kp}^t \geq 0, k = 1, ... n, \qquad (13)$$

***Definition 3.4***

Let $G(v_1, v_2, ... v_n)$ be a graph and the set (5) be a parabolic PDE on G. If coefficients satisfy the set (13), then equation (5) is called the diffusion equation on graph G.

$$f_p^{t+1} = \sum_{v_k \in U(v_p)} c_{pk}^t f_k^t, p = 1,2, ... n, \sum_{v_p \in U(v_k)} c_{pk}^t = 1, c_{kp}^t \geq 0, k = 1, ... n \qquad (14)$$

Let us show that the sum $S^t = \sum_{k=1}^n f_k^t$ of values of the function on all points of G for a diffusion equation (14) according to initial values does not depend on t.

***Theorem 3.2***

In the diffusion equation (14) on a graph $G(v_1, v_2, ... v_n)$, the sum $S^t = \sum_{k=1}^n f_k^t$ of values of the solution $f_p^t$ on all points of G according to initial values does not depend on t.

*Proof*

Consider the diffusion equation in the form (6).
$S^{t+1} = \sum_{p=1}^n f_p^{t+1} = \sum_{p=1}^n \sum_{k=1}^n c_{pk}^t f_k^t = \sum_{k=1}^n \sum_{p=1}^n c_{pk}^t f_k^t = \sum_{k=1}^n f_k^t \sum_{p=1}^n c_{pk}^t$. According to (13), $\sum_{p=1}^n c_{pk}^t = 1, k = 1, ... n$. Therefore, $S^{t+1} = \sum_{k=1}^n f_k^t = S^t$. Hence, $S^t = S^0$. The proof is complete. □

It is easy to see that equation (14) is stable.

***Theorem 3.3***

A diffusion equation (14) on a grah $G(v_1, v_2, ... v_n)$ is stable according to initial values, i.e., $\|f^t\| \leq \|f^0\|$, if for any t.

*Proof*

Consider the diffusion equation in the form (6).

$\|f^{t+1}\| = \sum_{p=1}^n |f_p^{t+1}| = \sum_{p=1}^n |\sum_{k=1}^n c_{pk}^t f_k^t| \leq \sum_{p=1}^n \sum_{k=1}^n |c_{pk}^t||f_k^t| = \sum_{k=1}^n \sum_{p=1}^n |c_{pk}^t||f_k^t| = \sum_{k=1}^n |f_k^t| \sum_{p=1}^n |c_{pk}^t|.$

According to (13), $\sum_{p=1}^n |c_{pk}^t| = \sum_{p=1}^n c_{pk}^t = 1, k = 1, ... n$. Therefore, $\|f^{t+1}\| \leq \sum_{k=1}^n |f_k^t| = \|f^t\|$. Hence, $\|f^t\| \leq \|f^0\|$. The proof is complete. □

At rather large times t, the form of a solution will be determined by a limit form of a matrix $C^t = \prod_{s=0}^t C(s)$. It is of some interest to clarify a behavior of matrix $C^t$ as $t \to \infty$. Call this limit as a final matrix $C^\infty$. The final matrix $C^\infty$ converts initial values $f_k^0$ into final values $f_k^\infty$. In the diffusion equation (14) in the matrix form (8), the square (n x n)-matrix $C^t$ is a stochastic matrix (or, more correct, a transposed stochastic matrix) which properties are well known (see [1] and [12]). Such matrices are used in Markov processes. Let's consider some applications of such approach for a solution of the initial value problem for the diffusion equation on a graph $G(v_1, v_2, ... v_n)$. Consider the case when matrix C(t) is not time dependent, $C(t) = C = \{c_{pk}\}$. Then the initial value problem for the diffusion equation on a graph $G(v_1, v_2, ... v_n)$ is the set (14) when

$$C(t) = C = \{c_{pk}\}, c_{pk} \geq 0, p, k = 1, ... n \qquad (15)$$

If matrix C is indecomposable and primitive, then it has a simple maximum eigenvalue 1, and there are no other complex eigenvalues which modulus are equal one (see [1] and [12] ). Matrix $C^t$ converges to a limit stochastic matrix $C^\infty$ as t→∞, which can be presented in the following form:

$$C^\infty = \begin{bmatrix} c_1 & \cdot & c_1 \\ \cdot & \cdot & \cdot \\ c_n & \cdot & c_n \end{bmatrix} = \lim_{t \to \infty}(C(t) \times C(t-1) ... \times C(0)), \sum_{k=1}^{n} c_k = 1, \forall c_k > 0. \qquad (16)$$

**Theorem 3.4**

If matrix $C = \{c_{pk}\}$ in the diffusion equation (14) is indecomposable and primitive, then the final solution $f^\infty$ of (14) as t→∞ at any initial values is stationary, not time-dependent (in each point of a graph, a value of function $f^\infty$ is constant, does not depend on time t), is determined only by the sum of values of function f in all points of the graph G and has the form:

$$f^\infty = C^\infty f^0 \qquad (17)$$

$$f^\infty = S \begin{bmatrix} c_1 \\ \cdot \\ c_n \end{bmatrix}, C^\infty = \begin{bmatrix} c_1 & \cdot & c_1 \\ \cdot & \cdot & \cdot \\ c_n & \cdot & c_n \end{bmatrix}, f^0 = \begin{bmatrix} f_1^0 \\ \cdot \\ f_n^0 \end{bmatrix}, \sum_{k=1}^{n} f_k^0 = S, \sum_{k=1}^{n} c_k = 1, \forall c_k > 0 \qquad (18)$$

*Proof*

$C^\infty = \lim_{t \to \infty}((C(t) \times ... C(0))$. According to [1] and [12], $C^\infty = \begin{bmatrix} c_1 & \cdot & c_1 \\ \cdot & \cdot & \cdot \\ c_n & \cdot & c_n \end{bmatrix}$. It is easy to check directly that $C^\infty f^0 = f^\infty$ and $Cf^\infty = f^\infty$. Therefore, $f^\infty$ the final solution of (14). The proof is complete.

Clearly, the solution $f^\infty$ depends only on S, but not on concrete distribution of values of function $f^0$ on points of graph G in the initial moment t = 0. Besides, any function of the form $f = df^\infty$ is an eigenvector of $C = C^0 = C^t$, and a stationary solution of the equation (13) because $Cf = CC(\infty)f = C(\infty)C = f$. That is

$$f_p = \sum_{v_k \in G} c_{pk} f_k \qquad (19)$$

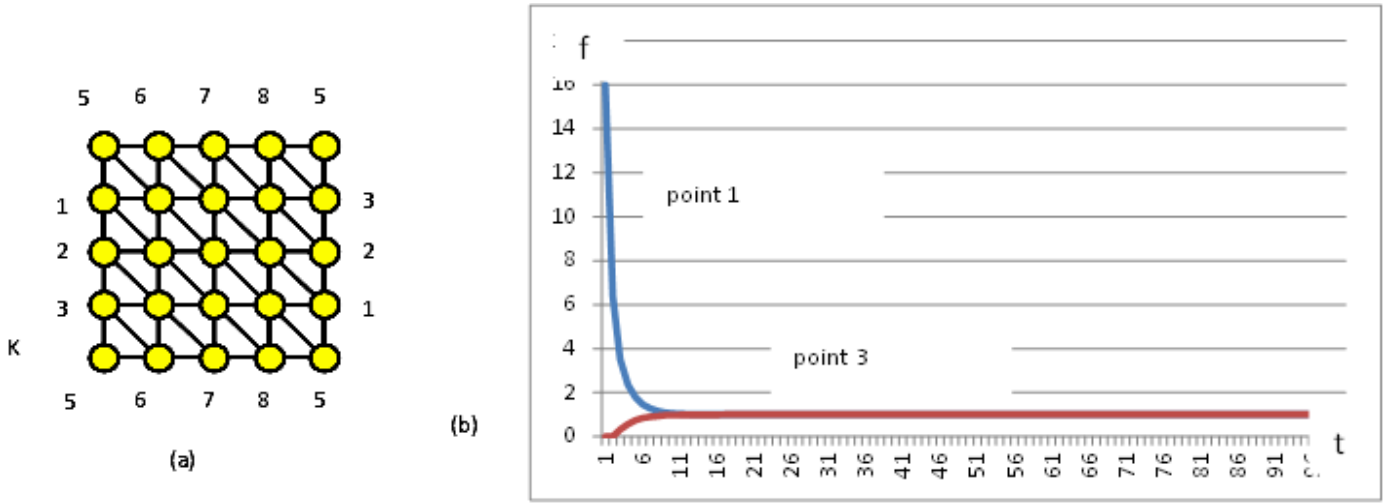

**Fig. 7.** (a) K is a digital Klein bottle containing 16 points. (b) The solution profile at point 1 and 3 on tne Klein bottle K, t=0, 1,... 100, $f_1^0$=16, $f_k^0$=0, k≠1.

This equation can be considered as an analog of an elliptical equation on a graph.

The accuracy of the solution of equation (5) can be increase by several ways. The first way is to tune functions $c_{pk}^t$ in (5) so that the accuracy is higher. This way is not the best because equation (5) must fit the given PDE and cannot be used for other equations. The other way is to use more points in the domain which is a graph (digital space) G. For example, an edge (uv) in G can be replaced by a point z using contractible transformations, and the obtained graph (space) $(G - (uv)) \cup z$ is homeomorphic to G (see [8]). In fig. 2, digital 2-sphere $S^2_{min}$ containing six points can be replaced by $S^2_3$ containing twelve points.

## 4. Numerical Solutions of a Diffusion Equation

Consider several examples of the numerical solutions of the diffusion equation (14) on different digital spaces.

$$f_p^{t+1} = \sum_{v_k \in U(v_p)} c_{pk} f_k^t = \sum_{k=1}^n c_{pk} f_k^t, p = 1,2,\ldots n, \quad (20)$$

$$\sum_{p=1}^n c_{pk} = 1, c_{pk} \geq 0, k, p = 1,2 \ldots n$$

### 4.1. Digital 2-dimensional Klein Bottle K. The Heat Equation on K

A Klein bottle has attracted attention of researcher in many fields. We mention here, among the others, physics [13], where a considerable interest has emerged in studying lattice models on non-orientable surfaces as new challenging unsolved lattice-statistical problems and as a realization and testing of predictions of the conformal field theory.

A digital Klein bottle K depicted in fig. 3 and fig. 7(a) consists of sixteen points. The rim $O(v_k)$ of every point $v_k$ is a digital 1-sphere containing six points. K is a homogeneous digital space containing the minimal number of points. The number of points can be increased by using contractible transformation. Topological properties of a digital 2D Klein bottle are similar to topological properties of its continuous counterpart. For example, the Euler characteristic and the homology groups of a continuous and a digital Klein bottle are the same ([10] and [11]).

Define coefficients $c_{pk}$ in the following way. If points $v_p$ and $v_k$ are adjacent then $c_{pk}$=0.1; $c_{kk}$=0.4, k=1,…16. Initial values are given as $f_1^0$=16, $f_k^0$=0, k=2,…16. In fig. 7(b), numerical solutions of the of the initial value problem for (20) are plotted at time t=1,…100. Two lines give the numerical solutions of the of the initial value problem for (20) at point 1 and point 3 (see fig. 7(a)). It follows from calculations that both lines converge to the same limit, $\lim_{t \to \infty} f_k^t = 1, k = 1,\ldots 16,$. The physical interpretation of (20) on K is the heat equation and the numerical solution shows the temperature at points of K.

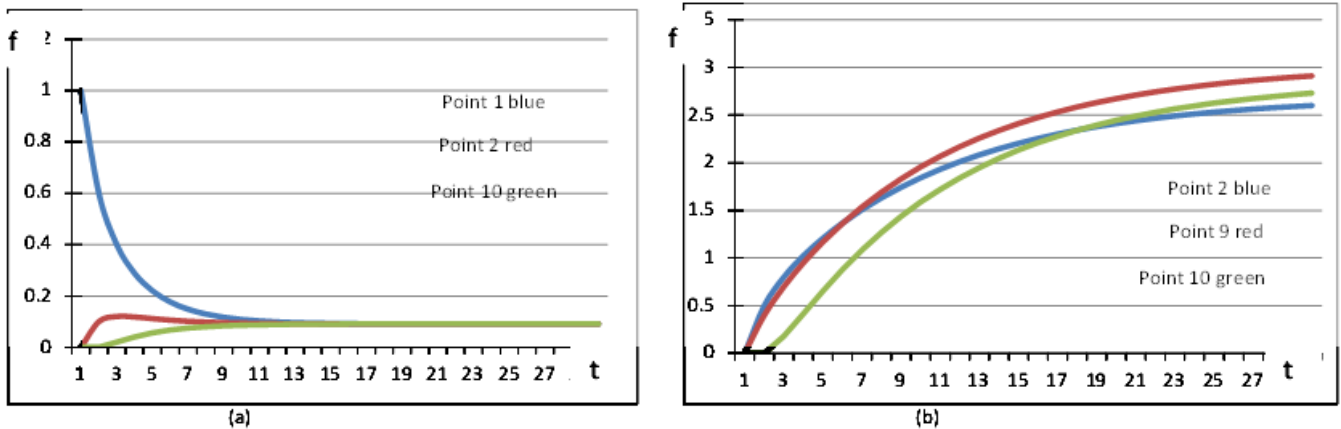

**Fig. 8.** Numerical solutions on the projective plane P. (a) The solution profile of the initial value problem at points 1, 2 and 10 on the projective plane P, t=0, 1,… 30, $f_1^0$=1, $f_k^0$=0, k≠1. (b) The solution profile of the boundary value problem at points 2, 8 and 11 on the Moebius strip M, t=0, 1,… 30, $f_1^t$=$f_{11}^t$=4.

### 4.2. Digital 2D Projective Plane P. The Diffusion Equation on P

Fig. 2 and 8 show a digital 2-dimensional projective plane P. P is a non-homogeneous digital space containing eleven points. The rim $O(v_k)$ of every point $v_k$ is a digital 1-sphere. Topological properties of a digital and a continuous 2D projective plane are similar. It was shown in [10] and [11] that the Euler characteristic and the homology groups of a continuous and a digital projective plane are the same. It is easy to check directly that a digital 2D projective plane without a point is homotopy equivalent to a digital 1D sphere as it is for a continuous projective plane.

Define coefficients $c_{pk}$ in the following way; if $v_k$ and $v_p$ are adjacent then $c_{kp} = c_{pk} = 0.1$, $c_{pp} = 1 - \sum_{k=1, k \neq p}^{11} c_{kp}$. Consider the numerical solutions of the initial value problem for (20). Initial values are given as $f_1^0 = 11$, $f_k^0 = 0$, k=2,…10. The profiles of the solution at points 1, 2 and 10 are plotted in fig. 8(a), where t=1,…30. It follows from direct calculations that all three lines converge to b=1. The physical meaning of these solutions is consistent with the distribution of particles on P caused by Brownian motion.

Consider the boundary value problem for (20). The boundary conditions are defined by equations $f_1^t = 1$, $f_{11}^t = 4$. Let $f_k^0 = 0$, k=2,…10. The profiles of the solution of the boundary value problem for (20) at points 2, 9 and 10 for t=1,…30 are shown in fig. 8(b).

This is consistent with the physical interpretation of (20) as heat equation, and the solution is the temperature distribution at points on P.

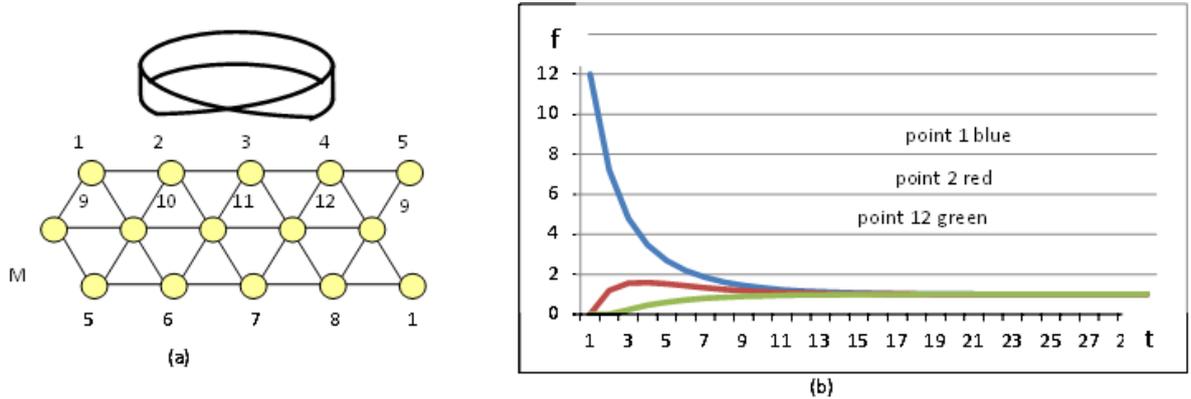

**Fig. 9.** *(a) The Moebius strip M is formed by interior points 9, 10, 11, 12 and boundary points 1-8. (b) The solution profile of the initial value problem at points 1, 2 and 12 on the Moebius strip M, t=0, 1,… 30, $f_1^0 = 12$, $f_k^0 = 0$, k≠1.*

### 4.3. Digital 2D Moebius Strip M. The Heat Equation on M

The Moebius strip M depicted in fig. 9(a) consists of twelve points. This is the minimal possible number of points. Points 9, 10, 11 and 12 are interior points, points 1-8 are boundary points which form a one-dimensional sphere (circle).

Assume that coefficients $c_{ik}$ do not depend on t; if $c_{ik} \neq 0$, i≠k, then $c_{ik} = c_{ki} = 0.1$, $c_{pp} = 0.4$, p=9,…12, $c_{pp} = 0.6$, p=1,…8. Initial values are given as $f_1^0 = 12$, $f_k^0 = 0$, k=2,…12. The profiles of the solution of (20) at points 1, 2 and 12 are shown in fig. 9(b), where t=0,…30. One may note that the difference between function values at different points dissipates with large passage of time. Direct computations show that $\lim_{t \to \infty} f_k^t = 1$, $k = 1, \ldots 12$. This is consistent with the physical interpretation of (20) as heat flow on M.

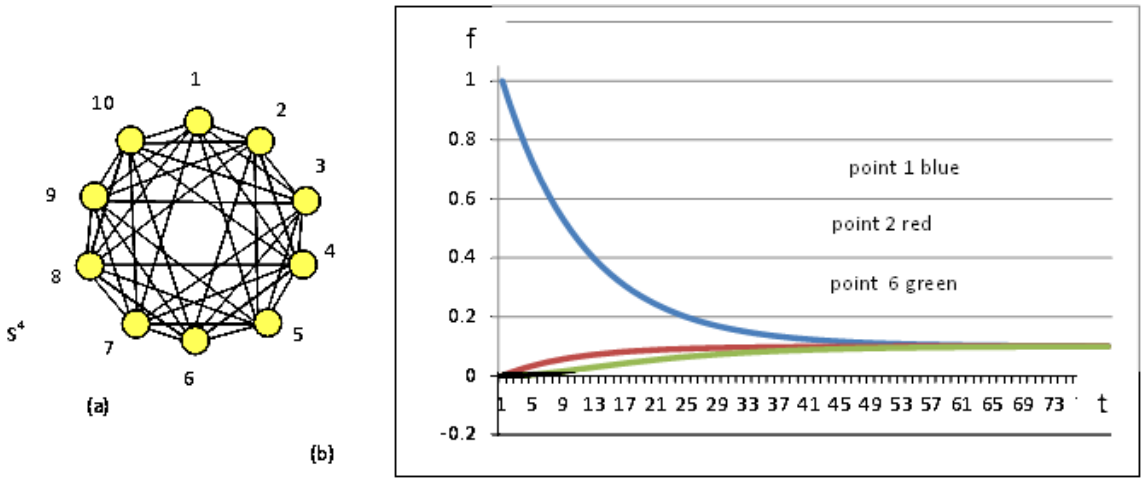

**Fig. 10.** *(a) $S^4$ is a digital 4-dimensional sphere with ten point. (b) The solution profile of the initial value problem at point 1, 2 and 6 on $S^4$, t=0, 1,… 80, $f_1^0 = 1$, $f_k^0 = 0$, k≠1.*

### 4.4. Digital 4-dimensional Sphere $S^4$. The Heat Equation on $S^4$

Consider a digital 4-dimensional -sphere $S^4$ with ten points depicted in fig. 4 and fig. 10(a). The rim $O(v_k)$ of every point $v_k$ is a digital 3-sphere containing eight points and depicted in fig. 4. $S^4$ is a homogeneous digital space containing the minimal number of points. The number of points can be increased by using contractible transformation. Topological properties of a digital 4D sphere are similar to topological properties of a continuous 4D sphere.

Consider the numerical solutions of the initial value problem for (20). Let $c_{ik}=c_{ki}=0.01$ if points $v_i$ and $v_k$ are adjacent, and $c_{pp}=0.92$ for p=1,…10. Initial values are given as $f^0_1=1$, $f^0_k=0$, k=2,…10. The results of the solution of the initial value problem (20) at points 1, 2 and 6 for t=0,…80 are displayed in fig. 10(b), which illustrates the time behavior of the values of the function f. In the physical heat interpretation, this is the temperature distribution on $S^4$. It follows from direct computations that $\lim_{t\to\infty} f_k^t = 0.1$, $k = 1,...10$.

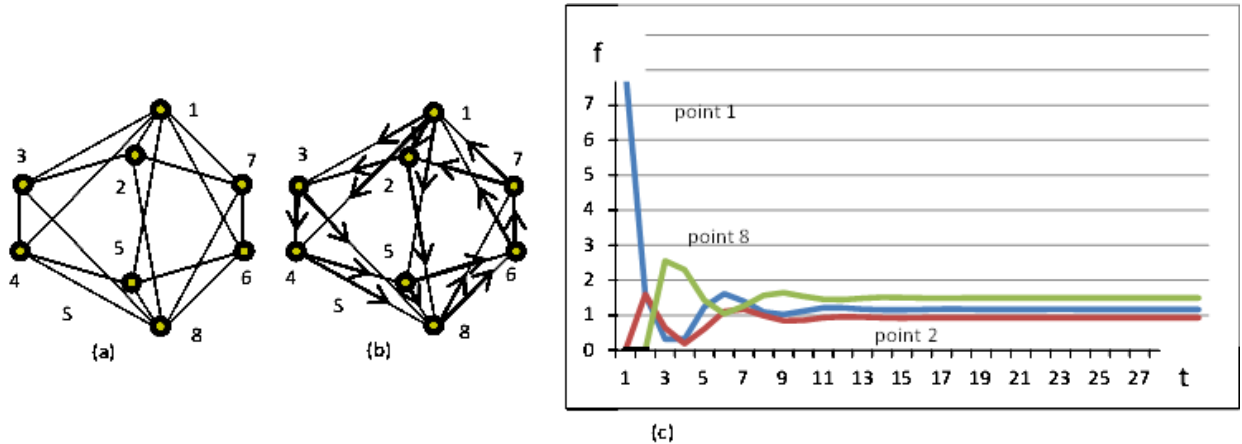

**Fig. 11.** *(a) A digital 2D sphere S. (b) A directed network N on S. (c) The solution profile of the initial value problem at point 1, 2 and 8 on $S^4$, t=1,… 30, $f_1^0=8$, $f_k^0=0$, k=2,… 8.*

### *4.5. Digital 2D Sphere S. A Directed Network*

Consider a digital 2-dimensional sphere S with eight points depicted in fig. 11(a). The rim $O(v_k)$ of every point $v_k$ is a digital 1-sphere. Define coefficients $c_{pk}$ in equation (20) in the following way; if $(v_iv_k) \in S$ then $c_{ik} > 0, c_{ki} = 0$ (or $c_{ik} = 0, c_{ki} > 0$), $c_{pp} = \sum_{k=1}^{8} c_{kp}$, p=1,…8. In this case, equation (20) forms a directed network N on S shown in fig. 11(b). Edge $(v_iv_k)$ in S can be replaced with a directed edge (or arc) $(v_i \to v_k)$ from $v_i$ to $v_k$. Dynamical system on such a network can be applied to a study of the movement of blood through the system of vessels or to an analysis of the of road traffic, ets. Let

$c_{12} = c_{13} = c_{14} = c_{15} = c_{11} = c_{22} = c_{33} = c_{44} = c_{55} = c_{66} = c_{77} = c_{88} = 0.2, c_{21} = c_{28} = c_{34} = c_{38} = c_{45} = c_{48} = c_{56} = c_{58} = c_{67} = c_{61} = c_{72} = c_{71} = c_{86} = c_{87} = 0.4$.

All other coefficients are equal to zero. Consider the numerical solutions of the initial value problem for (20). Initial values are given as $f_1^0=8, f_k^0=0$, k=2,…8. The results of the solution of (20) at points 1, 2 and 8 for t=1,…30 are displayed in fig. 11(c), which illustrates the time behavior of the values of the function. Dynamical system on such a network can be applied to a study of the movement of blood through the system of vessels or to an analysis of the of road traffic, ets.

## 6. Conclusion

A numerical explicit method for approximating the solution of differential parabolic equations on graphs, digital spaces, digital n-dimensional manifolds, n>0, and networks is presented and some properties of this method are investigated. This method is mathematically correct in terms of digital topology in the comparison with other methods used before. Solutions of the equation on a digital Klein bottle, a digital projective plane, a digital 4D sphere, a digital Moebius strip and a directed network are presented.